\documentclass[11pt]{article}
\usepackage{epsfig}
\usepackage{cite}

\begin{document}

\setcounter{page}{1}

\begin{center}
\begin{Large}
Measurement of the double longitudinal spin asymmetry in
inclusive jet production in polarized p+p collisions at $\sqrt{s} = 200$ GeV
\end{Large}

\vspace*{0.2in}
J. Kiryluk  for the STAR Collaboration \\
\vspace*{0.1cm}
{\footnotesize{\em{Massachusetts Institute of Technology 
\\ 77 Massachusetts Ave., Cambridge MA  02139-4307, USA}}
} \\

\end{center}

\vspace*{0.1in}

\begin{abstract}
We present preliminary results for the first measurements of the double longitudinal 
spin asymmetry $A_{LL}$  in inclusive jet production at mid-rapidity 
in polarized proton-proton collisions at $\sqrt{s} = 200\,\mathrm{GeV}$.
The data amount to $\sim 0.5$ pb$^{-1}$ collected at RHIC in 2003 and 2004 
with beam polarizations up to 45\%.
The jet transverse energies are in the range of $5 < p_T < 17\,\mathrm{GeV/c}$.
The data are consistent with theoretical evaluations using deep-inelastic 
scattering parametrizations for gluon polarization in the nucleon, 
and tend to disfavor large positive values of gluon polarization.
\end{abstract}

\vspace*{0.1in}

 One of the main objectives of the STAR (Solenoid Tracker At RHIC) spin physics program
is the precise determination of the polarized gluon distribution
in the nucleon by measurements of double longitudinal spin cross section asymmetries
$A_{LL}= \Delta \sigma/\bar{\sigma}=({\sigma^{++}-\sigma^{+-}})/({\sigma^{++}+\sigma^{+-}})$ 
in collisions of polarized protons at $\sqrt{s} = 200$ and 
$\sqrt{s} = 500$~$\mathrm{GeV}$~\cite{RHIC:Spin}.
The STAR setup provides charged particle tracking, particle identification, 
and electromagnetic calorimetry covering large acceptance \cite{nim:2003}.
These allow $A_{LL}$ measurements for various processes, including the production 
of inclusive jets and pions, of di-jets and di-hadrons, and of prompt-photon and 
jet coincidences.
The golden channel at STAR for a direct gluon polarization determination is the production
of a prompt-photon in coincidence with a jet, which is dominated by the gluon Compton 
process $q g \rightarrow \gamma q$. The polarized gluon distribution function $\Delta g(x)$ 
can be extracted from the measured asymmetry $A_{LL}$ over a wide and resolved kinematic range 
$0.01 < x < 0.3$. The gluon spin contribution to the proton spin is expected to be determined 
to a precision better than $0.5$~\cite{Gamma:1999} for projected future beam polarizations 
and luminosities at $\sqrt{s}=200$ and $\sqrt{s}=500$ GeV at RHIC~\cite{RHIC:Spin}.

While luminosity and polarization are being developed at RHIC, STAR has collected about $0.5\,\mathrm{pb}^{-1}$ 
of data in 2003 and 2004 at $\sqrt{s} = 200$ GeV.
The longitudinal average beam polarizations are $30-40$\% for these data.
The data allow an exploratory measurement of the asymmetry $A_{LL}$ in inclusive jet production, 
a process with a sizable cross section that is sensitive to the size of the gluon polarization in the proton:
\begin{small}
\begin{equation}
 A_{LL}^{\rm{jet}} =  \frac{
 \sum_{a,b}\int dx_a dx_b \times \Delta f_a(x_a,\mu^2) \times \Delta f_b(x_b,\mu^2)
\times \Delta \hat{\sigma}_{ab\rightarrow {\rm{jet}}X}(p_a,p_b,\alpha_S(\mu^2),p_\mathrm{T}^2/\mu^2) }
 {\sum_{a,b}\int dx_a dx_b \times f_a(x_a,\mu^2) \times  f_b(x_b,\mu^2)
 \times \hat{\sigma}_{ab\rightarrow {\rm{jet}}X}(p_a,p_b,\alpha_S(\mu^2),p_\mathrm{T}^2/\mu^2) }
\label{eq:all}
\end{equation}
\end{small}
\noindent
where $f_{a(b)}$ are the spin independent and $\Delta f_{a(b)}$ are the spin dependent parton distribution functions, 
and the subscripts $a,b$ refer to the initial partons in the hard interaction.
The hard partonic cross section is denoted as  $(\Delta) \hat{\sigma}_{ab\rightarrow {\rm{jet}}X}$.  
The parton four-momenta $p_{a(b)}$ are proportional to Bjorken $x_{a(b)}$ and the beam momenta.
The factorization and renormalization scales are assumed to be equal $\mu_F=\mu_R=\mu$ and $\mu^2 \sim p_T^2$.
Fragmentation functions enter a jet production measurement only indirectly, through reconstruction and trigger biases.
At leading order the cross section receives contributions from $gg\rightarrow gg$, $gq\rightarrow gq$  
and  $qq\rightarrow qq$.
Their relative contributions vary with $p_\mathrm{T}$ and are dominated by $gg$ and $qg$ 
scattering in the $p_\mathrm{T}$ range probed by STAR, $5 < p_T ({\rm{jet}}) < 17$ GeV/$c$.
The probed range of Bjorken $x$ of the parton is about $0.03-0.3$ at $\sqrt{s}=200$ GeV. 

Experimentally the double longitudinal spin asymmetry is defined as:
\begin{small}
\begin{equation}
A_{LL}^{\rm{jet}}= \frac{1}{P_1P_2} \left[ \frac{N^{++}-R N^{+-}}{ N^{++}+R N^{+-} } \right]
\end{equation}
\end{small}
\noindent
where $N_{++(+-)}$ are the inclusive jet yields for equal (opposite) spin orientations
of the protons, $R=L^{++}/L^{+-}$ is the ratio of luminosities for equal and opposite proton spin orientations,
and $P_{1(2)}$ are the proton beam polarization values.
RHIC polarimeters~\cite{CNI:2003} measure the degree of polarization and are ultimately 
expected to deliver an absolute measurement with 5\% uncertainty.
The polarization direction in the STAR interaction region is measured with Beam Beam Counters 
(BBC) \cite{BBC:2004}, as is the relative luminosity.

\begin{figure}
\begin{center}
  \includegraphics[height=.35\textheight]{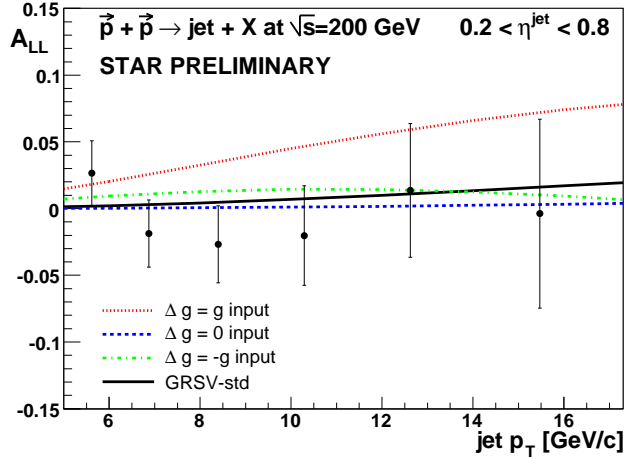}
  \caption{The double longitudinal spin asymmetry $A_{LL}$ in $\vec{p}+\vec{p} \rightarrow {\rm{jet}} +X$ 
at $\sqrt{s} = 200\,\mathrm{GeV}$ vs jet $p_T$. The curves show evaluations based on deep-inelastic 
scattering parametrizations of gluon polarization~\cite{Jager:2004}.}
\label{fig:all}
\end{center}
\end{figure}

Jets are reconstructed using a midpoint-cone algorithm~\cite{CDF:2000} with a cone size 
of $0.4$ that clusters charged tracks and electromagnetic energy deposits~\cite{Miller:2005}.
STAR's main tracker, the Time Projection Chamber (TPC), covers the pseudorapidity range 
$|\eta|<1.6$ and $2\pi$ in azimuth.
The Barrel ElectroMagnetic Calorimeter (BEMC) covered in 2003 and 2004 $0 < \eta < 1$ and $2\pi$ in azimuth.
The trigger was formed using coincidence BBC signals and a signal from a BEMC tower 
($\Delta \eta \times \Delta \phi = 0.05 \times 0.05$) above a transverse energy threshold of about 2.5\,GeV.
This trigger preferentially selects hard (quark) fragmentation and may thus bias the jet sample.
Selections in the analysis include the requirement of a vertex on the beam axis within $\pm 60$ cm 
of the nominal interaction point, a jet axis within a fiducial volume $0.2 < \eta_{\rm{jet}} < 0.8$,
and a sizable TPC contribution to the reconstructed jet energy to reject triggers caused by beam background,
jet $E_{\rm{EMC}}/E_{\rm{tot}} < 0.9$. The sample after selections consists of about $3\times10^{5}$ jets 
with transverse momenta of  $5 < p_T^{\rm{jet}} < 17$ GeV/$c$.
Fig.~\ref{fig:all} shows preliminary results for the double longitudinal spin asymmetry $A_{LL}$ in
inclusive jet production in polarized proton-proton collisions at $\sqrt{s} = 200\,\mathrm{GeV}$ 
from short data collection periods with longitudinally polarized proton beams in 2003 and 2004.
The indicated uncertainties are statistical.
We have considered systematic uncertainties from relative luminosity $R$, trigger bias, 
the possible contribution from residual non-longitudinal spin asymmetries, 
the contamination from beam background, 
and the beam polarizations. The total systematic uncertainty in the STAR measurement is about $0.01$, 
smaller than the statistical uncertainties. 
The RHIC beam polarization uncertainty amounts to an estimated $35(30) \%$ in scale for 2003(2004).  
Analyses with randomized spin patterns and other cross-checks including parity violating single-spin 
asymmetries show no evidence for beam bunch to bunch or fill to fill systematics.

The curves in Fig.~\ref{fig:all} show theoretical evaluations of $A_{LL}^{\rm{jet}}$ in inclusive 
jet production for different sets of polarized gluon distribution functions based on fits 
to deep-inelastic scattering data~\cite{Jager:2004,GRSV:2000,CTEQ6:2004}. 
They are based on a best fit - standard - GRSV polarized gluon distribution function (GRSV-std),
a vanishing polarized gluon distribution function  $\Delta g(x,Q^2_0) = 0 $ 
and a large positive or negative polarized gluon distribution function $\Delta g(x,Q^2_0) = \pm g(x,Q^2_0)$ 
at the input scale of the GRSV analysis $Q^2_0=0.6$\,GeV$^2$/c$^2$~\cite{GRSV:2000}.
The STAR jet asymmetries are consistent with three of these evaluations and
tend to be below the one that is based on the assumption 
that the gluons in the nucleon are maximally polarized.
Large and positive gluon polarization is thus disfavored, in agreement with 
Refs.~\cite{Compass:2005,Phenix:2005}.

In 2005 STAR recorded an integrated luminosity of 3pb$^{-1}$ with average beam polarizations 
of 45\% and a new trigger that required a transverse energy deposit larger than 7\,GeV within a BEMC patch 
($\Delta \eta \times \Delta \phi = 1.0 \times \pi/3 $).
This trigger is designed to be more efficient and is less biased.
The 2005 data analysis is in progress.

In summary, the first preliminary measurements of the double longitudinal spin asymmetries in mid-rapidity 
inclusive jet production in polarized proton-proton interactions at $\sqrt{s}=200 \mathrm{GeV}$ are presented.
The results are consistent with evaluations based on deep-inelastic scattering parametrizations 
for the gluon polarization in the nucleon, and tend to disfavor large positive values of gluon polarization.

\begin{thebibliography}{9}

\bibitem{RHIC:Spin}
G.~Bunce, N. Saito, J. Soffer, and W. Vogelsang, Ann.~Rev.~Nucl.~Part.~Sci. {\bf{50}} (2000) 525.

\bibitem{nim:2003}
Special Issue: RHIC and Its Detectors, Nucl.~Instrum.~Meth.{\bf{A499}} (2003).

\bibitem{Gamma:1999}
 L.~Bland, hep-ex/9907058, published in EPIC99 Workshop Proceedings, Bloomington, USA.

\bibitem{CNI:2003}
O.~Jinnouchi et al, AIP Conf.~Proc.~{\bf{675}} (2003) 424.

\bibitem{BBC:2004}
J.~Kiryluk et al, hep-ex/0501072, published in Spin 2004 Conference Proceedings, Trieste, Italy.


\bibitem{CDF:2000}
G.~Blazey et al,hep-ex/0005012,published in Batavia 1999,QCD and weak boson physics in RunII.

\bibitem{Miller:2005} M.~Miller, for the STAR Collaboration, these proceedings (2005).

\bibitem{Jager:2004}
B.~J${\rm{\ddot{a}}}$ger, M.~Stratmann and W.~Vogelsang, Phys.~Rev.~{\bf{D70}} (2004) 034010.

\bibitem{GRSV:2000}
M.~Gl${\rm{\ddot{u}}}$ck, E.~Reya, M.~Stratmann and W.~Vogelsang, Phys.~Rev.~{\bf{D63}} (2001) 094005.

\bibitem{CTEQ6:2004}
S.~Kretzer, H.L.~Lai, F.I.~Olness and W.K.~Tung, Phys.~Rev.~{\bf{D69}} (2004) 114005. 

\bibitem{Compass:2005} J.~Nassalski for the Compass Collaboration, these proceedings (2005).

\bibitem{Phenix:2005} K.~Boyle for the PHENIX Collaboration, these proceedings (2005).

\end{thebibliography}
\end{document}